\documentstyle[twoside,fleqn,espcrc2]{article}

\def\beq{\begin{equation}}
\def\eeq{\end{equation}}
\def\eql#1{\label{#1}}
\def\eqr#1{(\ref {#1})}

\def\Tr{\mbox{Tr}}

\newcommand{\AmS}{{\protect\the\textfont2
  A\kern-.1667em\lower.5ex\hbox{M}\kern-.125emS}}

\title{Field transformation and Monte Carlo simulations}

\author{M. Pepe\address{Dipartimento di Fisica, Universit\`a di Milano
and INFN, Sezione di Milano, Italy}}

\begin{document}

\begin{abstract}
A new method to compute observables at many values of the parameters $\lambda$
for a model with lattice action ${\cal{S}}(\phi, \lambda)$ is  described.
After fixing a reference set $\lambda^r$ of parameters, a single 
simulation is carried out by using a ``reference action''
${\cal{S}} (\phi^r, \lambda^r)$ to generate configurations
of the field $\phi^r$. 
Then a suitable analytic transformation is performed from the
configurations of $\phi^r$ to the ones corresponding to the action 
${\cal{S}} (\phi, \lambda)$. 
Such a transformation allows to obtain the observables  
for values of the parameters $\lambda$ close to $\lambda^r$.
I present studies on the reliability of the algorithm in the case 
of the $\phi^4$ model in 2 dimensions.
\end{abstract}

\maketitle

\section{INTRODUCTION}

Usually in Monte Carlo (MC) simulations observables are measured 
at the values of the Lagrangian parameters at 
which the run is carried out. However, it is often necessary to 
evaluate observables at many different values of the
parameters,
for instance, in problems requiring a fine tuning.
This poses dramatic limitations to the numerical studies.

The reweighting technique $\cite{RW}$ has long been a way out of this
problem.
Let us consider a single MC run performed according to a reference action 
${\cal{S}} [\phi , \lambda ^r]$ with 
given values $\lambda ^r$ for the Lagrangian parameters. Field
configurations are generated with probability 
$e ^{-{\cal{S}} [\phi , \lambda ^r]}$ and expectation values 
for observables are computed averaging over these configurations. Since the 
equilibrium distribution is known, physical 
quantities can be estimated for different parameters without performing a new simulation. 
The expectation value of an observable 
for the set of parameters $\lambda$  can be calculated as 
a reweighted average 
from the configurations 
generated by the MC simulation performed at the set $\lambda^r$. 
Moreover, the field configurations generated according to 
$e^{-{\cal{S}} [\phi ,\lambda ^r]}$ must be a good 
sampling for the distribution 
$e^{-{\cal{S}} [\phi ,\lambda ]}$. In the limit of an infinite number of 
configurations, $e^{-{\cal{S}} [\phi ,\lambda ]}$ is always correctly 
reconstructed but, as this is not the 
real case, there is a systematic distortion of the equilibrium
distribution. 
To minimize this effect, the parameter set $\lambda$ should not differ much
from the reference set $\lambda ^r$. 

Recently, an alternative technique $\cite{FT}$ has been proposed 
to compute 
observables at various values of the couplings and masses by a single MC 
simulation. This method is based on a "field transformation" from a  
reference field $\phi ^r$, distributed with probability proportional to 
$e^{-{\cal{S}} [\phi^r ,\lambda ^r]}$, to a field $\phi$ distributed 
according to $e^{-{\cal{S}} [\phi ,\lambda ]}$. 
The transformation is defined by the equation

\beq\eql{ft}
{\cal{D}} \phi\; e^{-{\cal{S}} [\phi ,\lambda]}
=
{\cal{D}} \phi^r e^{-{\cal{S}} [\phi^r ,\lambda^r]}
\eeq 

Thus, an importance sampling for the field $\phi^r$ can be transformed
into  an  importance sampling for the field $\phi$.
I will explore this technique by using a simplified
version of \eqr{ft}. More precisely, I consider only the terms of
the action depending on single sites.
This simplification implies that it is 
necessary to reweight the data for observables with a suitable remainder 
action $\delta {\cal{S}} [\phi^r, \lambda, \lambda^r ]$.
Work is in progress to take into account the full action. 

Here I study how reliable is the field transformation technique compared 
with the reweighting method. The model
I consider is the lattice $\phi ^4$ scalar field theory in two dimensions. 

\section{THE CASE STUDIED: THE TWO-DIMENSIONAL $\phi^4$ MODEL}

I consider the $2-d$ lattice scalar $\phi^4$ field theory 
with periodic boundary conditions and $L^2$ sites.
 
The lattice action is 

\beq
{\cal{S}} [\phi,\lambda 
]
=-\frac {1}{2}
\sum _{n \mu} \phi _n (\phi _{n+\mu} +\phi _{n-\mu})
+\sum _n v^{\lambda} (\phi _n )  
\eeq

\noindent
where ( $\lambda \equiv (\lambda _2,\lambda _4)$ )

\beq
v^{\lambda} (\phi _n )=
\frac 12 (\lambda_2  +4 ) \phi _n ^2 +\frac 14 \lambda_4 \phi _n^4
\eeq

\noindent
The sum over $\mu$ is to be intended in the positive directions.

The expectation value for a monomial 
$P[\phi ]=\phi _{n_1}\; \phi _{n_2} \ldots \phi _{n_k}$
in the field  is defined to be 
 
\beq
<P[\phi ]>_\lambda\equiv
\frac 1{\cal{Z}}
\int \prod _n d\phi _n\; e^{-{\cal{S}} [\phi ,\lambda ]} P [\phi ]
\eeq

\noindent
where ${\cal{Z}}$ is the partition function.

A generic field transformation $\phi=\phi (\phi ^r)$ is defined by the
jacobian matrix: 
 
\beq\eql{ftg}
{\cal{J}}_{nm} = \frac {\partial \phi _n}{\partial \phi ^r_m}
\eeq

\noindent
Re-expressing the field polynomial $P[\phi]$ in terms of $\phi^r$, we
obtain a function 
${\cal{P}}[\phi ^r]$ of the reference  field. 
The same transformation in the measure of the functional integral yields
\beq
\prod _n d\phi _n\; e^{-{\cal{S}} [\phi ,\lambda ]}=
\prod _n d\phi ^r_n\; e^{-{\cal{S}} [\phi ^r,\lambda ^r]+
\delta {\cal{S}} [\phi ^r,\lambda ,\lambda ^r] }
\eeq

\noindent
where $\delta {\cal{S}}$, the "remainder action", is   
 
\beq\eql{rag}
\delta {\cal{S}} [\phi ^r,\lambda ,\lambda ^r]=
-{\cal{S}} [\phi ,\lambda] +{\cal{S}} [\phi ^r, \lambda^r] + \Tr\; \ln {\cal{J}}
\eeq

\noindent
Now the Green function assumes the following form
 
\beq\eql{gf}
<P[\phi ]>_{\lambda}=
\frac
{<{\cal{P}}[\phi ^r] \; e^{\delta {\cal{S}} [\phi ^r]}>_{\lambda^r}}
{<e^{\delta {\cal{S}} [\phi ^r]}>_{\lambda^r}}
\eeq

The remainder action vanishes if \eqr{ft} is exactly solved, otherwise
it  enters as
a reweighting term.
The integration of \eqr{ftg} for a pair 
$\lambda \equiv (\lambda _2,\lambda _4)$ allows
to compute ${\cal{P}}[\phi ^r]$ and 
$\delta {\cal{S}} [\phi ^r,\lambda ,\lambda ^r]$
 for each MC generated configuration
of the reference field  $\phi^r$. Then by averaging over the reference
field configurations,
one obtains the Green function $<P[\phi ]>_{\lambda}$ corresponding
to the Lagrangian with parameters $\lambda$. 

\subsection{Form of the Field Transformation}

The field transformation defined by \eqr{ft} is not easily
integrable. Thus, a simpler case has been considered. 
In the simulation, I have used the following diagonal jacobian:
 
\beq\eql{jused}
{\cal{J}}_{nm} = \frac {\partial \phi _n}{\partial \phi ^r_m}=
C \delta _{nm} \;
e^{v^{\lambda} (\phi _n ) -v^{\lambda^r} (\phi _n^r )}
\eeq

\noindent
where $C$ is a constant. 
The solution of the previous differential equation with the condition 
$\phi _n =0$ when $\phi _n^r =0$ is 
 
\beq\eql{eqint}
\int _0^{\phi_n} d\phi\; e^{-v^{\lambda} (\phi)}= 
C \int _0^{\phi_n^r} d\phi\; e^{-v^{\lambda^r} (\phi )}
\eeq

\noindent
The fields $\phi$ and $\phi^r$ are non-compact and the constant $C$ is 
determined 
requiring that $\phi_n \rightarrow \infty$ when 
$\phi_n^r \rightarrow \infty$. 


With the choice \eqr{jused}, it follows

\beq 
\Tr\; \ln\; {\cal{J}} =\sum_n [v^{\lambda} (\phi _n ) - v^{\lambda^r}
(\phi_n^r )]
\eeq
 
\noindent
and consequently a non vanishing  remainder action has to be used in
\eqr{gf}  to evaluate Green functions as
a ratio of  reweighted averages.
 


\subsection{Field Transformation and Reweighting}

In this section I report on the tests of confidence performed for the 
field transformation method. These results are compared with those 
obtained by the reweighting technique. 
This analysis extends the first studies of reliability carried 
out for the $\phi^4$ scalar field theory in 4 dimensions $\cite{FT}$.
I have examined the susceptibility 
and the expectation value of the action ${\cal{S}} [\phi , \lambda ]$. 

The MC simulation has been performed at the reference point 
$\lambda^r =(-.2,.23)$ and 66 pairs of parameters have
been considered: 28 pairs had $\lambda_2 =\lambda_2^r$ (group
1), while
the remaining 38 ones  had $\lambda_4 =\lambda_4^r$ (group 2). Moreover,
independent MC simulations have been carried out to check the results.

In the reference simulation 210000 uncorrelated data for
the action and 40000 uncorrelated data have been collected for the
susceptibility; the MC test runs have amounted to 40000 uncorrelated data.

The figures below show the results for the expectation value of the action.
\vskip .45cm
\begin{figure}[h]
\vspace{4.9cm}
\includegraphics{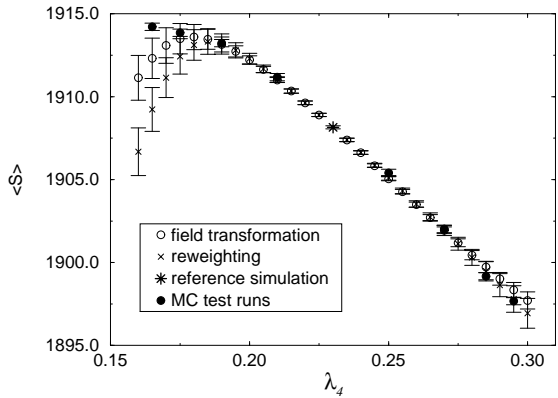}
\null
\vskip -1.3cm
\caption{Expectation value of the action for the group 1 pairs} 
\label{fig1}
\end{figure}
\vskip .45cm
\begin{figure}[h]
\vspace{4.cm}
\includegraphics{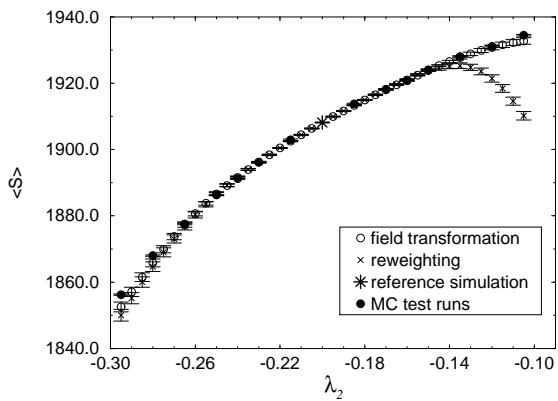}
\null
\vskip -1.3cm
\caption{Expectation value of the action for the group 2 pairs}
\label{fig2}
\end{figure}

\vskip -0.3cm
Both for the group 1 and the group 2 pairs, the reweighting technique,
the field transformation  method and the MC test runs
agree within the error bars, except
at small values of $\lambda_2$ and $\lambda_4$ where the
field transformation method performs better. 


The results for the susceptibility are analogous: 
the agreement between reweighting, field
transformation and MC test runs is within one standard
deviation provided  that the values of $\lambda_2$ and $\lambda_4$ are not
small and the critical region is not approached.


As a consequence of the choice \eqr{jused} for the field
transformation, the statistical error analysis is similar
to that performed for  the reweighting technique $\cite{errors}$.

\section{CONCLUSIONS}

The field transformation method is a generalization of the reweighting 
technique.
The simple case considered of a field transformation depending on
single sites 
does not provide a substantial improvement respect to reweighting.
Though the studies presented  are 
preliminary, the results obtained are similar to those given by the 
reweighting technique. However, including the full action in the field
transformation could yield a consistent improvement in the algorithm.
Work is in progress in this direction.
The aim is to modify the field
transformation in order to eliminate the need of the weight.

\end{document}